
\magnification=1200
\baselineskip=6mm
\nopagenumbers
\noindent
\centerline{\bf CONFORMAL INVARIANCE IN
CLASSICAL FIELD THEORY}
\vskip 2truecm
\centerline{D. R. Grigore\footnote*{e-mail: grigore@roifa.bitnet} }
\centerline{ Department of Theoretical Physics,
Institute of Atomic Physics}
\centerline{ Bucharest-Magurele, P.O. Box MG6,
Romania}
\vskip 2truecm
\centerline{ABSTRACT}
\vskip 1truecm
A geometric generalization of the first-order Lagrangian formalism
is used to analyse a conformal field theory for an
arbitrary primary field. We require that the
global conformal transformations are Noetherian symmetries and we prove
that the action functional can be taken strictly invariant
with respect to these transformations. In other words, there
does not exists a "Chern-Simons" type Lagrangian for a
conformally invariant Lagrangian theory.

\vskip1truecm
\vfil\eject

\footline={\hss\tenrm\folio\hss}

\pageno=1

{\bf 1. Introduction}
\vskip 1truecm
Conformal field theories continue to be a domain of active
research. (See for instance [1] and references quoted there).
For various reasons all investigations are done in the
framework of quantum field theory. The investigation of these
theories in the framework of classical field theory
seems to be absent from the literature. This can be explained,
as pointed out in [1] p. 190, because the local conformal
transformations cannot be defined everywhere in the complex plane;
the analiticity forces them to diverge somewhere.
So these type of transformations cannot be interpreted as
invariance transformations for a classical Lagrangian. However
one can consider only the global conformal transformations
(also called the homografic transformations) which can be
defined as bijective applications of the completed complex plane
$C \cup \{\infty\}$. We think that such an analysis is not without interest.

In particular, we are adressing the following question.
Let us consider a first-order Lagrangian theory for a
primary field. (This hypotesis is made for simplicity,
but in principle one can consider more than one primary field).
We impose the condition that the global conformal transformations
are Noetherian symmetries i.e. according to the usual definitions
(see e.g. [2], p. 16) the action functional is invariant up
to a trivial action ($\equiv$ an action giving trivial
Euler-Lagrange equations of motions).
We will be able to prove that one can redefine the action functional
such that it will be strictly invariant with respect to
these transformations. So, there are no Lagrangians of the
``Chern-Simons'' type for a (global) conformal field theory.

The main technical tool will be the geometric Lagrangian formalism
developped in [3]-[7] (see also [8]) for classical field
theory starting from original ideas of Poincar\'e, Cartan
and Lichnerowicz. The same technique was aplied by the
same author for studying gauge theories, gravitation theory, etc.
(see ref. [8] and references quoted there).

In Section 2 we present the general framework (more details
can be found in [8]). In Section 3
we particularize the results of Section 2 for a conformal
field theory. In Section 4 we derive the most general
Lagrangian theory (in the sense of Section 2) compatible
with conformal invariance.
\vskip1truecm
{\bf 2. General Theory}

2.1 Let $S$ be a differentiable manifold of dimension
$n+N$. The first order Lagrangian formalism is based on an
auxiliary object, namely the bundle of 1-jets of
$n$-dimensional submanifolds of $S$, denoted by
$J^{1}_{n}(S)$. This differentiable manifold is,
by definition:
$$J^{1}_{n}(S) \equiv \cup_{p \in S} J^{1}_{n}(S)_{p}$$
where $J^{1}_{n}(S)_{p}$ is the manifold of
$n$-dimensional linear subspaces of the tangent
space $T_{p}(S)$ at $S$ in the point $p \in S$.
This manifold is naturally fibered over $S$ and
we denote by $\pi$ the canonical projection.
Let us construct charts on $J^{1}_{n}(S)$ adapted
to this fibered structure. We first choose a local
coordinate system $(x^{\mu},\psi^{A})$ on the
open set $U \subseteq S$; here $\mu =1,...,n$
and $A=1,...,N$. Then on the open set $V \subseteq
\pi^{-1}(U)$, we shall choose the local coordinate
system $(x^{\mu},\psi^{A},{\chi^{A}}_{\mu})$,
defined as follows: if $(x^{\mu},\psi^{A})$ are the
coordinates of $p \in U$, then the
 $n$-dimensional plane in $T_{p_{0}}(S)$
corresponding to  $(x^{\mu},
\psi^{A},{\chi^{A}}_{\mu})$ is spanned
by the tangent vectors:
$${\delta \over \delta x^{\mu}} \equiv
{\partial \over \partial x^{\mu}} +
{\chi^{A}}_{\mu} {\partial \over \partial
\psi^{A}}.\eqno(2.1)$$

We will systematically use the summation convention
over the dummy indices.

By an {\it evolution space} we mean
any (open) subbundle $E$ of $J^{1}_{n}(S)$.

2.2 Let us define for a given evolution space $E$:
$$\Lambda_{LS} \equiv \{ \sigma \in \wedge^{n+1}(J^{1}_{n}(S))|
i_{Z_{1}}i_{Z_{2}} \sigma=0,\forall Z_{i},~s.~t.~ \pi_{*}Z_{i}=0,
i=1,2 \}.\eqno(2.2)$$

It is clear that any $\sigma \in \Lambda_{LS}$ can be written
in the local coordinates from above as follows:
$$\sigma =\varepsilon_{\mu_{1},...,\mu_{n}}
\sum_{k=0}^{n} {1 \over k!} C^{k}_{n}
\sigma^{\mu_{0},...,\mu_{k}}_{A_{0},...,A_{k}}
d{\chi^{A_{0}}}_{\mu_{0}} \wedge \delta \psi^{A_{1}}
\wedge...\wedge \delta \psi^{A_{k}} \wedge
dx^{\mu_{k+1}} \wedge...\wedge dx^{\mu_{n}}+$$
$$\varepsilon_{\mu_{1},...,\mu_{n}} \sum_{k=0}^{n}
{1 \over (k+1)!} C^{k}_{n}
\tau^{\mu_{1},...,\mu_{k}}_{A_{0},...,A_{k}}
\delta \psi^{A_{0}} \wedge...\wedge \delta \psi^{A_{k}}
\wedge dx^{\mu_{k+1}} \wedge...\wedge dx^{\mu_{n}}.\eqno(2.3)$$

Here $\varepsilon_{\mu_{1},...,\mu_{n}}$ is the
signature of the permutation $(1,...,n) \mapsto
(\mu_{1},...,\mu_{n})$, $\delta \psi^{A}$ is by definition:
$$\delta \psi^{A} \equiv d \psi^{A}- {\chi^{A}}_{\mu}
dx^{\mu}.\eqno(2.4)$$

We can suppose that the functions
$\sigma^{\mu_{0},...,\mu_{k}}_{A_{0},...,A_{k}}$ are
completely antisymmetric in the indices $\mu_{1},...,\mu_{k}$
and also in the indices $A_{1},...,A_{k}$, and the functions
$\tau^{\mu_{1},...,\mu_{k}}_{A_{0},...,A_{k}}$
are completely antisymmetric in the indices $\mu_{1},...,\mu_{k}$
and also in the indices $A_{0},...,A_{k}$.

It is remarkable that the following relation
has an intrinsic global meaning:
$$\sum_{i,j=1}^{k} (-1)^{i+j} \sigma^{\mu_{j},\mu_{1},...,
\hat{\mu_{j}},...,\mu_{k}}_{A_{i},A_{1},...,\hat{A_{i}},....,
A_{k}}=0.\eqno(2.5)$$
for $k = 1,...,n$.

This can be verified directly by
computing the transformation law for the functions
$\sigma^{...}_{...}$ with respect to a change of charts
on $E$ induced by a change of charts on $\pi(E) \subseteq S$.

More abstractely [6], one can prove this as follows.
One defines first the local operator
$K$ on $\Lambda_{LS}$ by:
$$K \sigma \equiv i_{{\delta \over \delta x^{\mu}}}
i_{{\partial \over \partial {\chi^{A}}_{\mu}}}
(\delta \psi^{A} \wedge \sigma).\eqno(2.6)$$
and proves that $K$
is in fact globally defined. Then one can show that (2.5)
is the local expression of the global relation:
$$K \sigma =0.\eqno(2.7)$$

We say that $\sigma \in \Lambda_{LS}$ is a
{\it Lagrange-Souriau form on E} if it verifies (2.7)
(or locally (2.5)) and is also closed:
$$d \sigma=0.\eqno(2.8)$$

In practical computations we will need the local
form of (2.8). By some work one arrives at the following
relations:
$${\partial \sigma^{\mu_{0},...,\mu_{k}}_{A_{0},...,A_{k}}
\over \partial {\chi^{A_{k+1}}}_{\mu_{k+1}}}-
\sigma^{\mu_{0},...,\mu_{k+1}}_{A_{0},...,A_{k+1}}-
(A_{0} \leftrightarrow A_{k+1},\mu_{0}
\leftrightarrow \mu_{k+1})=0.\eqno(2.9)$$
$${\delta \sigma^{\zeta,\rho,\mu_{1},...,\mu_{k}}_{B,A_{0},...,A_{k}}
\over \delta x^{\rho}}- \sum_{i=0}^{k} (-1)^{i}
{\partial \sigma^{\zeta,\mu_{1},...........,\mu_{k}}_{B,A_{0},...,
\hat {A_{i}},...,A_{k}} \over \partial \psi^{A_{i}}}+
{\partial \tau^{\mu_{1},...,\mu_{k}}_{A_{0},...,A_{k}} \over
\partial {\chi^{B}}_{\zeta}}-
\tau^{\mu_{1},...,\mu_{k},\zeta}_{A_{0},...,A_{k},B}=0.\eqno(2.10)$$
$${\delta \tau^{\mu_{0},...,\mu_{k}}_{A_{0},...,A_{k+1}}
\over \delta x^{\mu_{0}}}+\sum_{i=0}^{k+1} (-1)^{i}
{\partial \tau^{\mu_{1},............,\mu_{k}}_{A_{0},...,\hat{A_{i}},...
A_{k+1}} \over \partial \psi^{A_{i}}}=0.\eqno(2.11)$$
for $k = 0,...,n$. Here a hat means as usual omission.

We will call (2.5) and (2.9)-(2.11) the
{\it structure equations}.

A {\it Lagrangian system over S} is a couple $(E,\sigma)$
where $E \subseteq J^{1}_{n}(S)$ is some evolution
space over $S$ and $\sigma$ is a Lagrange-Souriau form on $E$.

2.3 The purpose of the Lagrangian formalism is to
describe {\it evolutions} i.e.
immersions $\Psi:M \rightarrow S$,
where $M$ is some $n$-dimensional manifold, usually
interpreted as the space-time manifold of the system.

Let us note that frequently, one supposes
that $S$ is fibered over $M$, but we do not need
this additional restriction in developing the general
formalism.
Let us denote by $\dot\Psi:M \rightarrow
J^{1}_{n}(S)$ the natural lift of $\Psi$.
If $(E,\sigma)$ is a Lagrangian system over $S$,
we say that $\Psi:M \rightarrow S$ verifies the
{\it Euler-Lagrange equations} if:
$$\dot\Psi^{*} i_{Z}\sigma=0.\eqno(2.12)$$
for any vector field $Z$ on $E$.

2.4 By a {\it symmetry} of the Euler-Lagrange equations
we understand a map $\phi \in Diff(S)$ such that if
$\Psi:M \rightarrow S$ is a solution of these
equations, then $\phi \circ \Psi$ is a solution
of these equations also.

It is easy to see that if $\phi \in Diff(S)$
is such that $\dot\phi$ leaves $E$ invariant and:
$$\dot\phi^{*} \sigma=\sigma.\eqno(2.13)$$
then it is a symmetry of the Euler-Lagrange equations (2.12).
We call the symmetries of this type {\it Noetherian symmetries}
for $(E,\sigma)$.

If a group $G$ act on $S$: $G \ni g \mapsto \phi_{g}
\in Diff(S)$ then we say that $G$ is a {\it group of Noetherian
symmetries} for $(E,\sigma)$ if for any $g \in G$,
$\phi_{g}$ is a Noetherian symmetry. In particular we have:
$$(\dot\phi_{g})^{*}\sigma=\sigma.\eqno(2.14)$$

It is considered of physical interest to solve the
following classification problem: given the manifold
$S$ with an action of some group $G$ on $S$,
find all Lagrangian systems $(E,\sigma)$ where
$E \subseteq J^{1}_{n}(S)$ is on open subset and
$G$ is a group of Noetherian symmetries for $(E,\sigma)$.
This goal will be achieved by solving simultaneously
(2.7), (2.8) and (2.14) in local coordinates and
then investigating the possibility of globalizing
the result.

2.5 Now we make the connection with the usual
Lagrangian formalism. We can consider that the
open set $V \subseteq \pi^{-1}(U)$ is simply
connected by choosing it small enough.

The first task is to exhibit somehow a Lagrangian.
To this purpose, we use the structure equations (2.9)-(2.11).
Using induction (from $k = n$ to $k = 0$) and applying
repeteadly the Poincar\'e lemma, one shows rather easily
that $\sigma^{...}_{...}$ and $\tau^{...}_{...}$ can be
written in the following form:
$$\sigma^{\mu_{0},...,\mu_{k}}_{A_{0},...,A_{k}}
={\partial L^{\mu_{1},...,\mu_{k}}_{A_{1},...,A_{k}}
\over \partial {\chi^{A_{0}}}_{\mu_{0}}}-
L^{\mu_{0},...,\mu_{k}}_{A_{0},...,A_{k}}.\eqno(2.15)$$
$$\tau^{\mu_{1},...,\mu_{k}}_{A_{0},...,A_{k}}=
\sum_{i=0}^{k} (-1)^{i} {\partial
L^{\mu_{1},...........,\mu_{k}}_
{A_{0},...,\hat{A_{i}},...,A_{k}}
\over \partial \psi^{A_{i}}} -
{\delta L^{\mu_{0},...,\mu_{k}}_
{A_{0},...,A_{k}} \over \delta x^{\mu_{0}}}.\eqno(2.16)$$
where $L^{\mu_{1},...,\mu_{k}}_{A_{1},...,A_{k}}$ are
(real) functions defined on $V$ and completely antisymmetric in the
indices $\mu_{1},...,\mu_{n}$ and also in the indices
$A_{1},...,A_{n}$. A more abstract way to prove this
is given in [6]. From (2.8) one has in $V$:
$$\sigma = d\theta.\eqno(2.17)$$
for some $n$-form $\theta$. Then one can show that
by eventually redefining $\theta$: $\theta \rightarrow
\theta + df$ one can exhibit it in the form:
$$\theta \equiv \varepsilon_{\mu_{1},...,\mu_{n}}
\sum_{k=0}^{n} {1 \over k!} C^{k}_{n} L^{\mu_{1},...,
\mu_{k}}_{A_{1},...,A_{k}} \delta \psi^{A_{1}} \wedge
...\wedge \delta \psi^{A_{k}} \wedge dx^{\mu_{k+1}}
\wedge ... \wedge dx^{\mu_{n}}.\eqno(2.18)$$

Then (2.15) and (2.16) follow from (2.3), (2.17) and (2.18).

Finally, using the structure equation (2.5) one gets a recurrence
relation for the functions
$L^{\mu_{1},...,\mu_{k}}_{A_{1},...,A_{k}}$
and easily shows that:
$$L^{\mu_{1},...,\mu_{k}}_{A_{1},...,A_{k}}
\equiv {1 \over k!} \sum_{\sigma \in P_{k}} (-1)^{|\sigma|}
{\partial^{k} L \over \partial {\chi^{A_{1}}}_{\mu_
{\sigma(1)}}...\partial {\chi^{A_{k}}}_{\mu_{\sigma(k)}}}.
\eqno(2.19)$$
($P_{k}$ is the permutation group of the numbers $1,...,k$) and
$|\sigma|$ is the signature of $\sigma$).
$L$ is called a {\it local Lagrangian}. If $\sigma$ is of
the form (2.3) with the coefficients $\sigma^{...}_{...}$
and $\tau^{...}_{...}$ given by (2.16), (2.17) and (2.20) then
we denote it by $\sigma_{L}$.
The expressions (2.18)-(2.19) are exactly the ones appearing in
[3]-[5]. So, we can conclude that the framework above
generalizes the scheme from [3]-[5] in the folowing sense: the
central object is now the Lagrange-Souriau form $\sigma$, not
the Poincar\'e-Cartan form $\theta$. The connection between them
is only local (see (2.17)) and $\sigma$ can be globally defined.

Now one can easily show that the local form of the Euler-Lagrange equations
(2.12) coincides with the usual one. Namely, one
chooses convenient local coordinates $(x^{\mu},\psi^{A})$
on an open set $U \subseteq S$, such that the evolution
$\Psi:M \rightarrow S$ will be
locally given by $x^{\mu} \mapsto (x^{\mu},\Psi^{A}(x))$.
Then $\dot\Psi:M \rightarrow J^{1}_{n}(S)$  is given by
 $\dot\Psi$ by $x^{\mu} \mapsto (x^{\mu},\Psi^{A}(x),
{\partial \Psi^{A} \over \partial x^{\mu}}(x))$ and
the equations (2.12) have the local expression:
$${\partial L \over \partial \psi^{A}}
\circ \dot\Psi- {\partial \over \partial x^{\mu}}
\left( {\partial L \over \partial
{\chi^{A}}_{\mu}} \circ \dot\Psi \right)=0.\eqno(2.20)$$
i.e. the usual Euler-Lagrange equations if one takes
$\sigma = \sigma_{L}$.

We also note the following result:
The Euler-Lagrange equations (2.12) are trivial
{\it iff} $\sigma = 0$. One can prove this fact elementary.
Indeed if $\sigma = 0$, then it is clear that the Euler-Lagrange
equations (2.12) are trivial. Conversely, suppose that (2.12)
are identities. Substituting (2.3) into (2.12) one obtaines that
$\sigma^{\mu_{0}\mu_{1}}_{A_{0}A_{1}} \equiv 0$
and
$\tau_{A_{0}} \equiv 0$.
Now one uses the structure equations (2.5) + (2.7)-(2.11) to
prove by induction that
$\sigma^{...}_{...} \equiv 0$
and
$\tau^{...}_{...} \equiv 0$,
i.e. $\sigma \equiv 0$. For another proof see [7].

Let us suppose  now for the moment that $\sigma$ is exact i.e.
verifies (2.17) on the whole $E$.
Then one can define the action functional by the formula:
$${\cal A}(\Psi) \equiv \int_{M} \dot\Psi^{*} \theta.\eqno(2.21)$$

One can show that $\Psi$ is a solution of
the Euler-Lagrange equations {\it iff} it is
an extremal of the action functional.

We say that the functional $A$ is a {\it trivial action}
if it gives trivial Euler-Lagrange equations. Then the
usual definition of Noetherian symmetries is encoded in
the relation:
$${\cal A}(\phi \circ \Psi) ={\cal A}(\Psi) + a~trivial~action\eqno(2.22)$$
(see [2]).

Now one can establish that in this case (2.22) is equivalent to
the definition (2.13) given in the general case.
To prove this assertion one proceeds as follows. First, one
notes that in this case one can write $\theta$ as $\theta_{L}$
for some Lagrangian $L$, and (2.21) becomes the usual expression:
$${\cal A}(\Psi) = \int_{M} L \circ \dot\Psi.\eqno(2.23)$$

Next, one plugs this expression into (2.22) and obtaines (taking
into account that $\Psi$ is arbitrary):
$$J_{\phi} L \circ \dot\phi = L + L_{0}.\eqno(2.24)$$

Here $J_{\phi}$ follows from the (eventual) volume change
induced by $\phi$ and $L_{0}$ is a trivial Lagrangian i.e. a
Lagrangian giving trivial Euler-Lagrange equations of motions.

Now it is a matter of computation to show that (2.24) is
equivalent to the following relation:
$$\dot\phi^{*} \theta_{L} = \theta_{L} + \theta_{L_{0}}.\eqno(2.25)$$

Applying the exterior derivative one gets:
$$\dot\phi^{*} \sigma_{L} = \sigma_{L} + \sigma_{L_{0}}.\eqno(2.26)$$

But, because the Euler-Lagrange equations for $L_{0}$ are
trivial, we have $\sigma_{L_{0}} = 0$ according to a remark
above, so we get (2.13) ($\sigma = \sigma_{L}$ in our case).
Conversely, if one has (2.13) then applying Poincar\'e lemma one
obtaines (2.25) with $\theta_{L_{0}}$ an exact $n$-form. But in
this case $\sigma_{L_{0}} = 0$ and the same remark above implies
that $L_{0}$ is a trivial Lagrangian. The proof is finished.

The equivalence between (2.13) and (2.22)
shows clearly why (2.13) is the most suitable
definition of Noetherian symmetries: it makes sense also in the
case when $\sigma$ is not exact (and (2.22) makes no sense).

We close this subsection with a comment about the necessity of
working with the rather complicated expression (2.3). It is more
or less obvious from what has been said above that the proof of
the equivalence between (2.13) and (2.22) relies heavily on the
specific structure of $\sigma$ given by (2.3). Moreover one can
show [5] that if one considers only a truncated $\sigma$ (for
instance one takes in (2.18) the sum from $0$ to $p < n$) then
one can save the implication (2.12) $\Longrightarrow$ (2.20) if $p
\geq 1$ but, in general, only the implication (2.13) $\Longrightarrow$
(2.22) stays true. In other words, one can consider in (2.3)
only the terms $k = 0,1,2$ and still has a geometrical way of
expressing the Euler-Lagrange equations (2.12), but the set of
transformations $\phi$ satisfying (2.13) is, in general,
strictly smaller that the whole group of Noetherian symmetries.
Only when one considers the whole sum these two sets are identical.

2.6 An important particular case of the
general framework presented in the Section 2.2 is the following one.
We say that a Lagrangian system $(E,\sigma)$ is of the
{\it Chern-Simons type} if:
$$\sigma \in \Lambda_{CS} \equiv
\{ \sigma \in \Lambda^{n+1}(E)\vert i_{Z} \sigma=0,
\forall Z,~s.~t.~ \pi_{*}Z=0 \}.\eqno(2.27)$$
(compare with (2.2)). The terminology is justified by the
following fact [10]: for a gauge theory such a
$\sigma$
follows from a Chern-Simons Lagrangian.

It is clear that in the local coordinates used up till now, such a
$\sigma$ is of the form (2.3) where one makes:
$\sigma^{...}_{...} \rightarrow 0$.

One can prove that in this case the (local) Lagrangian
can be chosen a polynomial in the variables ${\chi^{A}}_{\mu}$
of maximal degree $n$.
\vskip1truecm

{\bf 3. Conformal Field Theories}

3.1 We consider only the simplest case of a primary field.
In the general framework of Section 2, we take
$S=R^{2} \times R^{2}$ with coordinates $(x^{1},x^{2},\psi^{1},\psi^{2})$.
As it is well known, it is more convenient to use as independent
variables the following complex combinations:
$$z^{\mu} \equiv x^{1} + \mu i x^{2}.\eqno(3.1)$$
$$\psi^{\mu} \equiv \psi^{1} + \mu i \psi^{2}.\eqno(3.2)$$
for $\mu = +,-$. As we have said in the Introduction, we will
consider that in fact $z^{\mu}$ takes values in the completed
complex plane $C \cup \{\infty\}$.
We take for the evolution space the manifold
$E \equiv J^{1}_{2}(S)$ with coordinates
$(z^{\mu},\psi^{\nu},{\chi^{\nu}}_{\mu})$
We particularize now the expression (2.3) of $\sigma$ :
$$\sigma=2\varepsilon_{\mu_{1},\mu_{2}}
\sigma^{\mu_{0},\mu_{1}}_
{\nu_{0},\nu_{1}} d{\chi^{\nu_{0}}}_{\mu_{0}} \wedge
\delta \psi^{\nu_{1}} \wedge
\wedge dz^{\mu_{2}} + \varepsilon_{\mu_{1},\mu_{2}}
\sigma^{\mu_{0},\mu_{1},\mu_{2}}_{\nu_{0},\nu_{2},\nu_{2}}
d{\chi^{\nu_{0}}}_{\mu_{0}} \wedge
\delta \psi^{\nu_{1}} \wedge \delta \psi^{\nu_{2}} +$$
$$\varepsilon_{\mu_{1},\mu_{2}}
\tau_{\nu_{0}} \delta \psi^{\nu_{0}} \wedge
dz^{\mu_{1}} \wedge dz^{\mu_{2}} +
\varepsilon_{\mu_{1},\mu_{2}}
\tau^{\mu_{1}}_{\nu_{0},\nu_{1}} \delta \psi^{\nu_{0}}\wedge
\delta \psi^{\nu_{1}} \wedge
dz^{\mu_{2}}.\eqno(3.3)$$
where we have taken into account the fact that (2.5)
for $k = 1$ gives $\sigma^{\mu_{0}}_{\nu_{0}} = 0$.
Here:
$$\delta \psi^{\nu} \equiv d \psi^{\nu}-{\chi^{\nu}}_{\mu} dz^{\mu}.
\eqno(3.4)$$

We list now the structure equations. From (2.5) we have:
$$
\sigma^{\mu_{0},\mu_{1}}_{\nu_{0},\nu_{1}} -
\sigma^{\mu_{1},\mu_{0}}_{\nu_{0},\nu_{1}} -
\sigma^{\mu_{0},\mu_{1}}_{\nu_{1},\nu_{0}} +
\sigma^{\mu_{1},\mu_{0}}_{\nu_{1},\nu_{0}} = 0.\eqno(3.5)$$

{}From (2.9) we have:
$$\sigma^{\mu_{0},\mu_{1}}_{\nu_{0},\nu_{1}} =
\sigma^{\mu_{1},\mu_{0}}_{\nu_{1},\nu_{0}}.\eqno(3.6)$$
$${\partial \sigma^{\mu_{0},\mu_{1}}_{\nu_{0},\nu_{1}}
\over \partial {\chi^{\nu_{2}}}_{\mu_{2}}}-
\sigma^{\mu_{0},\mu_{1},\mu_{2}}_{\nu_{0},\nu_{1},\nu_{2}}-
(\mu_{0}\nu_{0} \leftrightarrow \mu_{2}\nu_{2})=0.\eqno(3.7)$$
$${\partial \sigma^{\mu_{0},\mu_{1},\mu_{2}}_{\nu_{0},\nu_{2},\nu_{2}}
\over \partial {\chi^{\nu_{3}}}_{\mu_{3}}} - (\mu_{0}\nu_{0}
\leftrightarrow \mu_{3}\nu_{3}) = 0.\eqno(3.8)$$

{}From (2.10) we have:
$${\delta \sigma^{\zeta,\rho}_{\omega,\nu_{0}} \over \delta z^{\rho}} +
{\partial \tau_{\nu_{0}} \over
\partial {\chi^{\omega}}_{\zeta}}-\tau^{
\zeta}_{\nu_{0},\omega}=0.\eqno(3.9)$$
$${\delta \sigma^{\zeta,\rho,\mu_{1}}_{\omega,\nu_{0},\nu_{1}}
\over \delta z^{\rho}} -
{\partial \sigma^{\zeta,\mu_{1}}_{\omega,\nu_{1}} \over
\partial \psi^{\nu_{0}}} +
{\partial \sigma^{\zeta,\mu_{1}}_{\omega,\nu_{0}} \over
\partial \psi^{\nu_{1}}} +
{\partial \tau^{\mu_{1}}_{\nu_{0},\nu_{1}} \over
\partial {\chi^{\omega}}_{\zeta}} = 0.\eqno(3.10)$$
$${\partial \sigma^{\zeta,\mu_{1},\mu_{2}}_
{\omega,\nu_{1},\nu_{2}} \over \partial \psi^{\nu_{0}}} -
{\partial \sigma^{\zeta,\mu_{1},\mu_{2}}_
{\omega,\nu_{0},\nu_{2}} \over \partial \psi^{\nu_{1}}} +
{\partial \sigma^{\zeta,\mu_{1},\mu_{2}}_
{\omega,\nu_{0},\nu_{1}} \over \partial \psi^{\nu_{2}}} = 0.\eqno(3.11)$$

Finally, from (2.11) we have:
$${\delta \tau^{\mu_{0}}_{\nu_{0},\nu_{1}}
\over \delta z^{\mu_{0}}}+
{\partial \tau_{\nu_{1}} \over \partial \psi^{\nu_{0}}} -
{\partial \tau_{\nu_{0}} \over \partial \psi^{\nu_{1}}} = 0.\eqno(3.12)$$
$${\partial \tau^{\mu_{1}}_{\nu_{1},\nu_{2}} \over
\partial \psi^{\nu_{0}}} -
{\partial \tau^{\mu_{1}}_{\nu_{0},\nu_{2}} \over
\partial \psi^{\nu_{1}}} +
{\partial \tau^{\mu_{1}}_{\nu_{0},\nu_{1}} \over
\partial \psi^{\nu_{2}}} = 0.\eqno(3.13)$$

Here:
$${\delta \over \delta z^{\mu}} \equiv {\partial \over
\partial z^{\mu}}+{\chi^{\nu}}_{\mu} {\partial \over
\partial \psi^{\nu}}.\eqno(3.14)$$

3.2 We now impose the invariance with respect to
global conformal transformations. Let $f^{+}$ and $f^{-}$ be
homografic functions of the variables $z^{+}$ and $z^{-}$
respectively. For consistency we must require that:
$$(f^{+}(z^{+}))^{*} = f^{-}(z^{-}).\eqno(3.15)$$

We denote by $f$ the map:
$$f(z^{+},z^{-}) \equiv (f^{+}(z^{+}),f^{-}(z^{-})).\eqno(3.16)$$

Then we consider the following transformation on $S$:
$$\phi_{f}(z^{\mu},\psi^{\nu}) = \left(f^{\mu}(z^{\mu}),
\prod_{\rho} \left[\dot f^{\rho}(z^{\rho})\right]^{m_{\nu\rho}}
\psi^{\nu}\right).\eqno(3.17)$$

Here $m_{+}$  and $m_{-}$ are two real numbers characteristic
of the primary field $\psi$ (the conformal weights).
By $\dot f$ we denote the first derivative of $f$.

We say that the Lagrangian system $(E,\sigma)$ is {\it globally conformal
invariant } if for any $f^{\mu}$ we have:
$$(\dot \phi_{f})^{*} \sigma = \sigma.\eqno(3.18)$$

Let us make the connection with the usual definition of
conformal invariance. We suppose for the moment that $\sigma
=d\theta$ and we define the action functional by (2.22).

If $f$ is as above then the action of the corresponding
conformal transformation on the set of immersions
$\Psi:M \rightarrow S$ (for a 2-dimensional manifold $M$) is:
$$\Phi_{f}(z^{\mu},\Psi^{\nu}(\cdot)) =
\left( f^{\mu}(z^{\mu}),\left[ {\partial f^{+} \over
\partial z^{+}}(\cdot)\right]^{m_{\nu}}\left[
{\partial f^{-} \over \partial z^{-}}(\cdot)\right]^{m_{-\nu}}
\Psi^{\nu} \circ f^{-1} (\cdot)\right).\eqno(3.19)$$
(see e.g. [1]).

Then (3.18) is equivalent to the usual definition of
conformal invariance:
$${\cal A}(\Phi_{f} \Psi) = {\cal A}(\Psi) + a~trivial~action.\eqno(3.20)$$

3.3 According to the strategy outlined in Section 2 we work
with the more general definition (3.18). It is convenient
to use this relation in an infinitesimal form. Namely,
one takes:
$$f^{\mu}(z^{\mu}) = z^{\mu} + \theta^{\mu}(z^{\mu}).\eqno(3.21)$$
with $\theta^{\mu}$ infinitesimally small, computes the
variation of $\sigma$ with respect to $\dot \phi_{f}$,
and takes into account the fact that $\theta^{+}$ and
$\theta^{-}$ are of the following form (see [1] p. 190):
$$\theta^{\mu}(z^{\mu}) = a_{-1}^{\mu} + a_{0}^{\mu} z^{\mu} +
a_{1}^{\mu} (z^{\mu})^{2}~~~(\mu = \pm)\eqno(3.22)$$
with $a_{-1}$, $a_{0}$ and $a_{1}$ arbitrary complex numbers.

We will give below the result of this elementary but tedious computation.
It is convenient to define the following differential operators on $E$:
$$D_{\rho} \equiv \sum_{\nu} m_{\nu\rho} \psi^{\nu}
{\partial \over \partial \psi^{\nu}} + \sum_{\mu,\nu}
(m_{\nu\rho} - \delta_{\mu\rho}) {\chi^{\nu}}_{\mu}
{\partial \over \partial {\chi^{\nu}}_{\mu}}.\eqno(3.23)$$
and:
$$D'_{\rho} \equiv \sum_{\nu} m_{\nu\rho} \psi^{\nu}
{\partial \over \partial {\chi^{\nu}}_{\rho}}.\eqno(3.24)$$

Then (3.18) is infinitesimally equivalent to the following set
of relations:
$${\partial \sigma^{\mu_{0},\mu_{1}}_{\nu_{0},\nu_{1}}
\over \partial z^{\rho}}=0.\eqno(3.25)$$
$$D_{\rho} \sigma^{\mu_{0},\mu_{1}}_{\nu_{0},\nu_{1}} +
\left(m_{\nu_{0}\rho} + m_{\nu_{1}\rho} - \delta^{\mu_{0}}_{\rho} +
\delta^{\mu_{1}}_{-\rho}\right) \sigma^{\mu_{0},\mu_{1}}_{\nu_{0},\nu_{1}}
= 0.\eqno(3.26)$$
$$D'_{\rho} \sigma^{\mu_{0},\mu_{1}}_{\nu_{0},\nu_{1}} = 0.\eqno(3.27)$$
$${\partial \sigma^{\mu_{0},\mu_{1},\mu_{2}}_{\nu_{0},\nu_{1},\nu_{2}}
\over \partial z^{\rho}} = 0.\eqno(3.28)$$
$$D_{\rho} \sigma^{\mu_{0},\mu_{1},\mu_{2}}_{\nu_{0},\nu_{1},\nu_{2}} +
\left(m_{\nu_{0}\rho} + m_{\nu_{1}\rho} + m_{\nu_{2}\rho}
 - \delta^{\mu_{0}}_{\rho}\right)
\sigma^{\mu_{0},\mu_{1},\mu_{2}}_{\nu_{0},\nu_{1},\nu_{2}}
= 0.\eqno(3.29)$$
$$D'_{\rho} \sigma^{\mu_{0},\mu_{1},\mu_{2}}_
{\nu_{0},\nu_{1},\nu_{2}} = 0.\eqno(3.30)$$
$${\partial \tau_{\nu_{0}}
\over \partial z^{\rho}}=0.\eqno(3.31)$$
$$D_{\rho} \tau_{\nu_{0}} + (m_{\nu_{0}\rho} + 1) \tau_{\nu_{0}}
= 0.\eqno(3.32)$$
$$D'_{\rho} \tau_{\nu_{0}} + \cdots = 0.\eqno(3.33)$$
$${\partial \tau^{\mu_{1}}_{\nu_{0},\nu_{1}}
\over \partial z^{\rho}}=0.\eqno(3.34)$$
$$D_{\rho}  \tau^{\mu_{1}}_{\nu_{0},\nu_{1}} +
\left(m_{\nu_{0}\rho} + m_{\nu_{1}\rho} +
 + \delta^{\mu_{1}}_{-\rho}\right) \tau^{\mu_{1}}_{\nu_{0},\nu_{1}}
= 0.\eqno(3.35)$$
$$D'_{\rho} \tau^{\mu_{1}}_{\nu_{0},\nu_{1}} + \cdots = 0.\eqno(3.36)$$
In (3.33) and (3.36) we mean by
$\cdots$ some algebraic expressions in $\sigma^{...}_{...}$
which will not be needed.

In the following section we will analyse separately
the cases $(m_{+})^{2} + (m_{-})^{2} \not= 0$ and $m_{+} = m_{-} = 0$
and we will prove that $\sigma$ can be exhibited in the form
$\sigma_{L} = d\theta_{L}$ such that the corresponding action functional
is strictly invariant with respect to $\dot \phi_{f}$ defined by (3.19) above.
The idea is to first analyse the functions $\sigma^{...}_{...}$
and then to eliminate them completely from the game and
ending up with a Chern-Simons Lagrangian theory. Of course,
the result anticipated above means that this last
contribution will be in fact trivial.
\vskip 2truecm
{\bf 4. The main theorem}

a) $(m_{+})^{2} + (m_{-})^{2} \not= 0$

4.1 In this case we use instead of the variables
${\chi^{\nu}}_{\mu}$ the new coordinates:
$$X_{\nu} \equiv \sum_{\nu} \mu m_{-\mu\nu}
\psi^{-\mu} {\chi^{\mu}}_{\nu}.\eqno(4.1)$$
and:
$$Y_{\nu} \equiv \sum_{\nu} m_{\mu\nu}
\psi^{-\mu} {\chi^{\mu}}_{\nu}.\eqno(4.2)$$

The Jacobian $D(X,Y)/D(\chi)$ is non-singular
outside the hypersurface $\psi^{+}\psi^{-} = 0$.

Then it easily follows that (3.27) and (3.30) means that
$\sigma^{...}_{...}$ do not depend on the variables $Y_{\nu}$.
So, taking into account the independence of $z^{\rho}$
(see (3.25) and (3.28)) it follows that $\sigma^{...}_{...}$
are functions only of $\psi$ and $X_{\nu}$ i.e.
$$\sigma^{\mu_{0},\mu_{1}}_{\nu_{0},\nu_{1}} =
s^{\mu_{0},\mu_{1}}_{\nu_{0},\nu_{1}} \circ \Phi.\eqno(4.3)$$
and:
$$\sigma^{\mu_{0},\mu_{1},\mu_{2}}_{\nu_{0},\nu_{1},\nu_{2}} =
s^{\mu_{0},\mu_{1},\mu_{2}}_{\nu_{0},\nu_{1},\nu_{2}} \circ
\Phi.\eqno(4.4)$$
where:
$$\Phi(z,\psi,\chi) \equiv (\psi,m_{-}\psi^{-} {\chi^{+}}_{+} -
m_{+}\psi^{+} {\chi^{-}}_{+},m_{+}\psi^{-} {\chi^{+}}_{-} -
m_{-}\psi^{+} {\chi^{-}}_{-}).\eqno(4.5)$$

4.2 Next one rephrases (3.26) and (3.29) in terms of $s^{...}_{...}$.
If we define:
$$\hat{D}_{\rho} \equiv \sum_{\nu} m_{\nu\rho} \psi^{\nu}
{\partial \over \partial \psi^{\nu}} + \sum_{\nu}
\left( m_{+} + m_{-} -\delta_{\nu\rho}\right) X_{\nu}
{\partial \over \partial X^{\nu}}.\eqno(4.6)$$
then we get respectively:
$$\hat{D}_{\rho} s^{\mu_{0},\mu_{1}}_{\nu_{0},\nu_{1}} +
\left(m_{\nu_{0}\rho} + m_{\nu_{1}\rho} - \delta^{\mu_{0}}_{\rho} +
\delta^{\mu_{1}}_{-\rho}\right) s^{\mu_{0},\mu_{1}}_{\nu_{0},\nu_{1}}
= 0.\eqno(4.7)$$
and:
$$\hat{D}_{\rho} s^{\mu_{0},\mu_{1},\mu_{2}}_{\nu_{0},\nu_{1},\nu_{2}} +
\left(m_{\nu_{0}\rho} + m_{\nu_{1}\rho} + m_{\nu_{2}\rho}
 - \delta^{\mu_{0}}_{\rho}\right)
s^{\mu_{0},\mu_{1},\mu_{2}}_{\nu_{0},\nu_{1},\nu_{2}}
= 0.\eqno(4.8)$$

4.3 We pursue with the analysis of the functions $s^{...}_{...}$
translating everything in terms of a Lagrangian $L_{0}$
depending only on the variables $\psi$ and $X$. To this purpose
we start with the structure equations (3.8). Inserting (4.4)
in this relation and applying Frobenius theorem it is clear
that one can find a system of functions $L^{\mu_{1}\mu_{2}}_
{\nu_{1}\nu_{2}}$ depending on $\psi$ amd $X$ , with
antisymmetry with respect to the transposition of  the indices
$\mu$ and also with respect to the transposition of the indices
$\nu$ and such that:
$$\sigma^{\mu_{0},\mu_{1},\mu_{2}}_{\nu_{0},\nu_{1},\nu_{2}} =
{\partial \over \partial {\chi^{\nu_{0}}}_{\mu_{0}}}
\left( L^{\mu_{1}\mu_{2}}_
{\nu_{1}\nu_{2}} \circ \Phi\right).\eqno(4.9)$$

Inserting (4.4) and (4.9) into (4.8) we get an equation of the
same type for $L^{\mu_{1}\mu_{2}}_
{\nu_{1}\nu_{2}}$:
$$\hat{D}_{\rho} L^{\mu_{1}\mu_{2}}_{\nu_{1}\nu_{2}} +
\left(m_{\nu_{1}\rho} + m_{\nu_{2}\rho} \right)
L^{\mu_{1}\mu_{2}}_{\nu_{1}\nu_{2}} = \cdots.\eqno(4.10)$$
where by $\cdots$ we mean
some $\psi$-dependent functions.

We iterate the procedure. Inserting (4.9) into (3.7) we get
as before that $\sigma^{\mu_{0},\mu_{1}}_{\nu_{0},\nu_{1}}$
are of the form:
$$\sigma^{\mu_{0},\mu_{1}}_{\nu_{0},\nu_{1}} =
{\partial \over \partial {\chi^{\nu_{0}}}_{\mu_{0}}} \left( L^{\mu_{1}}_
{\nu_{1}} \circ \Phi\right) - L^{\mu_{0}\mu_{1}}_{\nu_{0}\nu_{1}}
.\eqno(4.11)$$
with $L^{\mu_{1}}_{\nu_{1}}$ depending only on $\psi$ and $X$.
Also we use (4.3) and (4.11) into (4.7) and end up with:
$$\hat{D}_{\rho} L^{\mu_{1}}_{\nu_{1}} +
\left( m_{\nu_{1}\rho} + \delta^{\mu_{1}}_{-\rho}\right)
L^{\mu_{1}}_{\nu_{1}} =
\cdots.\eqno(4.12)$$
where by $\cdots$ we mean some expression depending only
on $\psi$ and $X$ which is polynomial in $X$ of maximal
degree 1.

Finally, we insert (4.11) into (3.6) and find out that
there exists a $(\psi,X)$-dependent function $L_{0}$
such that:
$$L^{\mu}_{\nu} = {\partial \over \partial {\chi^{\nu}}_{\mu}}
\left( L_{0} \circ \Phi\right).\eqno(4.13)$$

The derivation of (4.9), (4.11) and (4.13) is a process of
the same type as the process leading to the equations (2.15).

If we insert (4.13) into (4.12) we can prove that in fact
the right hand side of (4.12) does not depend on $X$
and we have:
$$\hat{D}_{\rho} L_{0} + L_{0} = F^{\mu}_{\rho} X_{\mu} +
 F_{\rho}.\eqno(4.14)$$
for some $\psi$-dependent functions $F^{\mu}_{\rho}$ and $F_{\rho}$.

Now the structure equation (3.5) expresses
$L^{\mu_{1}\mu_{2}}_{\nu_{1}\nu_{2}}$ also in terms of $L_{0}$:
$$L^{\mu_{1}\mu_{2}}_{\nu_{1}\nu_{2}} \circ \Phi =
{1 \over 2} {\partial^{2} \over \partial
{\chi^{\nu_{1}}}_{\mu_{1}} \partial {\chi^{\nu_{2}}}_{\mu_{2}}}
\left( L_{0} \circ \Phi\right) - (\mu_{1} \leftrightarrow
\mu_{2}).\eqno(4.15)$$

Again (4.13) and (4.15) should be compared with (2.19).

4.4 We try to eliminate completely the functions $\sigma^{...}_{...}$.
As a byproduct we will show that $L_{0}$ (which is not unique)
can be chosen such that in (4.14) the right hand side
is in fact $X$-independent.

To this purpose we define:
$$L \equiv L_{0} \circ \Phi.\eqno(4.16a)$$
$$\sigma_{CS} \equiv \sigma -
\sigma_{L}.\eqno(4.16b)$$

It is clear that $\sigma_{CS}$ is of the Chern-Simons type
and also that (3.18) is equivalent to:
$$(\dot\phi_{f})^{*} \sigma_{L} -\sigma_{L} +
(\dot\phi_{f})^{*} \sigma_{CS} -\sigma_{CS} = 0.\eqno(4.17)$$

It is better to compute separatedly the first two terms and the
last two terms in (4.17), using of course an infinitesimal
transformation. For the first two terms it is convenient
to compute first the variation of $\theta_{L}$ and then to
use $\sigma = d\theta_{L}$. For the last two terms in (4.17)
we make $\sigma^{...}_{...} \rightarrow 0$ and
$\tau^{...}_{...} \rightarrow (\tau_{CS})^{...}_{...}$
in the computations of Section 2. If we denote:
$$P_{\rho} \equiv F^{\mu}_{\rho} X_{\mu} \circ \Phi +
F_{\rho}.\eqno(4.18)$$
then it is easy to prove that (4.17) gives:
$${\partial (\tau_{CS})_{\nu_{0}} \over
\partial z^{\rho}}= 0.\eqno(4.19)$$
$$D_{\rho}(\tau_{CS})_{\nu_{0}} +
\left( m_{\nu_{0}\rho} + 1\right) (\tau_{CS})_{\nu_{0}} +
{\partial P_{\rho} \over \partial \psi^{\nu_{0}}} = 0.\eqno(4.20)$$
$$D'_{\rho} (\tau_{CS})_{\nu_{0}} +
{\partial P_{\rho} \over \partial {\chi^{\nu_{0}}}_{\rho}} = 0.\eqno(4.21)$$
and:
$${\partial (\tau_{CS})^{\mu_{1}}_{\nu_{0},\nu_{1}} \over
\partial z^{\rho}}= 0.\eqno(4.22)$$
$$D_{\rho}(\tau_{CS})^{\mu_{1}}_{\nu_{0},\nu_{1}} +
\left( m_{\nu_{0}\rho} + m_{\nu_{1}\rho} + \delta^{\mu_{1}}_{-\rho}
\right) (\tau_{CS})^{\mu_{1}}_{\nu_{0},\nu_{1}} -
\left({\partial^{2} P_{\rho} \over \partial {\chi^{\nu_{0}}}_{\rho}
\partial \psi^{\nu_{1}}} - (\nu_{0} \leftrightarrow \nu_{1})
\right) = 0.\eqno(4.23)$$
$$D'_{\rho} (\tau_{CS})^{\mu_{1}}_{\nu_{0},\nu_{1}} = 0.\eqno(4.24)$$
(compare with (3.31)-(3.35)).

We still have the structure equations for $(\tau_{CS})^{...}_{...}$.
These equations can be obtained from (3.9), (3.10), (3.12) and (3.13) making
$\sigma^{...}_{...} \rightarrow 0$ and
$\tau^{...}_{...} \rightarrow (\tau_{CS})^{...}_{...}$; we get:
$$
{\partial (\tau_{CS})_{\nu_{0}} \over
\partial {\chi^{\omega}}_{\zeta}}-(\tau_{CS})^{
\zeta}_{\nu_{0},\omega}=0.\eqno(4.25)$$
$$
{\partial (\tau_{CS})^{\mu_{1}}_{\nu_{0},\nu_{1}} \over
\partial {\chi^{\omega}}_{\zeta}} = 0.\eqno(4.26)$$
$$
{\partial (\tau_{CS})_{\nu_{1}} \over \partial \psi^{\nu_{0}}} -
{\partial (\tau_{CS})_{\nu_{0}} \over
\partial \psi^{\nu_{1}}} = 0.\eqno(4.27)$$
$${\partial (\tau_{CS})^{\mu_{1}}_{\nu_{1},\nu_{2}} \over
\partial \psi^{\nu_{0}}} -
{\partial (\tau_{CS})^{\mu_{1}}_{\nu_{0},\nu_{2}} \over
\partial \psi^{\nu_{1}}} +
{\partial (\tau_{CS})^{\mu_{1}}_{\nu_{0},\nu_{1}} \over
\partial \psi^{\nu_{2}}} = 0.\eqno(4.28)$$

It is rather easy to analyse (4.19)-(4.28). From (4.22)
and (4.26) it follows that $(\tau_{CS})^{\mu_{1}}_{\nu_{0},\nu_{1}}$
is only a $\psi$-dependent function. Then (4.19) and (4.25) give:
$$(\tau_{CS})_{\nu_{0}}(\psi,\chi) = t_{\nu_{0}}(\psi) +
(\tau_{CS})^{\mu_{1}}_{\nu_{0},\nu_{1}}(\psi)
{\chi^{\nu_{1}}}_{\mu_{1}}.\eqno(4.29)$$

Substituting this expression into (4.24) it follows that in fact:
$$(\tau_{CS})^{\mu_{1}}_{\nu_{0},\nu_{1}} = 0.\eqno(4.30)$$

Then (4.29) tells us that $(\tau_{CS})_{\nu_{0}}$ is only a
$\psi$-dependent function. Inserting this information into (4.21) we get:
$${\partial P_{\rho} \over \partial {\chi^{\nu_{0}}}_{\rho}} = 0$$
so, taking into account (4.18) it follows that in fact:
$$F^{\mu}_{\rho} = 0$$

This relation gives the following form of (4.14):
$$\hat{D}_{\rho} L_{0} + L_{0} =
 F_{\rho}.\eqno(4.31)$$
(We recall that $F_{\rho}$ is $\psi$-independent).

Because $(\tau_{CS})_{\nu_{0}}$ is only a
$\psi$-dependent function, the relation (4.27) gives us:
$$(\tau_{CS})_{\nu_{0}} = {\partial l \over \partial \psi^{\nu_{0}}}
.\eqno(4.32)$$
for some $\psi$-dependent function $l$. Now, it is quite easy to
see that if we redefine: $L_{0} \rightarrow L_{0} - l$, then we have:
$$\sigma =\sigma_{L}.\eqno(4.33)$$
and (4.31) stays true if we redefine conveniently the functions $F_{\rho}$.

But now the invariance conditions are equivalent to:
$${\partial F_{\rho} \over \partial \psi^{\nu_{0}}} = 0.\eqno(4.34)$$
(see (4.20)) i.e. $F_{\rho}$ are some constants.
{}From (4.31) it follows immediately that in this case
we must have $F_{+} = F_{-} \equiv F$.

So, by redefining $L_{0} \rightarrow L_{0} - F$, we
do not afect (4.33), but instead of (4.31) we have:
$$\hat{D}_{\rho} L_{0} + L_{0} = 0.\eqno(4.35)$$

4.5 The equations (4.33) and (4.35) provides us with the
most general solution in the case a).

In fact, the analysis can be pushed a little bit further
if we note that (4.35) is the infinitesimal form of:
$$\lambda_{+}\lambda_{-} L_{0}(\lambda_{+}^{m_{+}}\psi^{+},
\lambda_{-}^{m_{-}}\psi^{-},\lambda_{+}^{m_{+}+m_{-}-1}
\lambda_{-}^{m_{+}+m_{-}}X_{+},\lambda_{+}^{m_{+}+m_{-}}
\lambda_{-}^{m_{+}+m_{-}-1}X_{-}) = L_{0}(\psi,X).\eqno(4.36)$$
for any $\lambda_{+},\lambda_{-} \in R_{+}$.
(This in turn is equivalent to the invariance of the corresponding
action with respect to $\Phi_{f}$ defined by (3.19)).

If we take $\lambda_{\mu} = X_{\mu}$ for $\mu = +,-$ then
(4.36) tells us that $L_{0}$ is of the following form:
$$L_{0}(\psi,X) = X_{+}X_{-} l(X_{+}^{m_{+}}\psi^{+},
X_{-}^{m_{-}}\psi^{-},(X_{+}X_{-})^{m_{+}+m_{-}}).\eqno(4.37)$$

b) $m_{+} = m_{-} = 0$

In the same way as at a) we can prove that $\sigma$ can
be exhibited in the form (4.33) where $L$ verifies:
$$\sum_{\nu} {\chi^{\nu}}_{\mu} {\partial L \over
\partial {\chi^{\nu}}_{\mu}} + L = 0.\eqno(4.38)$$
which is the infinitesimal form of:
$$\lambda_{+}\lambda_{-} L(\psi,\lambda_{+}
{\chi^{\nu}}_{+},\lambda_{-} {\chi^{\nu}}_{-}) =
L(\psi,{\chi^{\nu}}_{+}, {\chi^{\nu}}_{-}).\eqno(4.39)$$
for any $\lambda_{\mu} \in R_{+}$.
(Again this expresses the invariance of the corresponding
action with respect to $\Phi_{f}$).

So, we have:

{\bf Theorem} {\it Any conformally invariant first-order Lagrangian
theory for a primary field is of the form $(E,d\theta_{L})$
where $L$ is such that the corresponding action functional
is strictly invariant with respect to $\Phi_{f}$:
$$A(\Phi_{f}\Psi) = A(\Psi).\eqno(4.40)$$

In particular, if $m_{+}m_{-} \not= 0$ then $L$ is determined
by $L = L_{0} \circ \Phi$ with $L_{0}$ of the form (4.37)
and if $m_{+} = m_{-} = 0$ then $L$ is determined by the
homogeneity property (4.39).}

{\bf Remark} It is quite plausible that a result of the same type
stays true for a more general case of a Lagrangian theory
depending on two or more primary fields.
\vskip1truecm

{\bf References}
\item{1.} Y. Grandati, ``\'Elements d'introduction \`a
l'invariance conforme'', Annales de Physique Fr. 17 (1992) 159-300
\item{2.} N. N. Bogoliubov, D. V. Shirkov, ``Introduction to
the Theory Quantized Fields'', third edition, John Wiley, 1980
\item{3.} D. Krupka, ``A Map Associated to the Lepagean Forms on
the Calculus of Variations in Fibered Manifolds'', Czech. Math.
Journ. 27 (1977) 114-118
\item{4.} D.Betounes, ``Extension of the Classical Cartan Form'',
Phys. Rev. D 29 (1984) 599-606
\item{5.} D.Betounes, ``Differential Geometrical Aspects of the
Cartan Form: Symmetry Theory'', J. Math. Phys. 28 (1987) 2347-2353
\item{6.} H.Rund, ``A Cartan Form for the Field Theory of
Charath\'eodory in the Calculus of Variation of Multiple Integral'',
Lect. Notes in Pure and Appl. Math. 100 (1985) 455-469
\item{7.} D.R.Grigore and O.T.Popp, `` On the Lagrange-Souriau
Form in Classical Field Theory'', submitted for publication
\item{8.} D.R.Grigore ``A Generalized Lagrangian Formalism
in Particle Mechanics and Classical Field Theory'',
Fortsch. der Phys. 41 (1993) no. 7
\bye